\documentclass[
    final            
  4apaper]
  {aipproc}

\layoutstyle{8x11single}

\newcommand\beq{\begin{equation}}
\newcommand\eeq{\end{equation}}
\newcommand\beqa{\begin{eqnarray}}
\newcommand\eeqa{\end{eqnarray}}

\newcommand{\text}{\mathrm}

\newcommand{\NS}{\text{NS}}

\newcommand{\dd}{\text{d}}

\begin{document}

\title{Longitudinal Viscous Flow in Granular Gases}


\author{Andr\'es Santos}{
  address={Departamento de F\'{\i}sica, Universidad de
Extremadura, E-06071 Badajoz, Spain} }

\begin{abstract}
The   flow characterized by a linear longitudinal velocity field
$u_x(x,t)=a(t)x$, where $a(t)={a_0}/({1+a_0t})$, a uniform  density
$n(t)\propto a(t)$,  and a uniform temperature $T(t)$ is analyzed
for dilute granular gases by means of a BGK-like model kinetic
equation in $d$ dimensions. For a given value of the coefficient of
normal restitution $\alpha$, the relevant control parameter of the
problem is the reduced deformation rate $a^*(t)=a(t)/\nu(t)$ (which
plays the role of the Knudsen number), where $\nu(t)\propto
n(t)\sqrt{T(t)}$ is an effective collision frequency. The relevant
response parameter is a nonlinear viscosity function $\eta^*(a^*)$
defined from the difference between the normal stress $P_{xx}(t)$
and the hydrostatic pressure $p(t)=n(t)T(t)$. The main results of
the paper are: (a) an exact first-order ordinary differential
equation for $\eta^*(a^*)$ is derived  from the kinetic model; (b) a
recursion relation for the coefficients of the Chapman--Enskog
expansion of $\eta^*(a^*)$ in powers of $a^*$ is obtained; (c)  the
Chapman--Enskog expansion is shown to diverge for elastic collisions
($\alpha=1$) and converge for inelastic collisions ($\alpha<1$), in
the latter case with a radius of convergence that increases with
inelasticity; (d) a simple approximate analytical solution for
$\eta^*(a^*)$, hardly distinguishable from the numerical solution of
the differential equation, is constructed.
\end{abstract}

\maketitle


\section{Introduction}
A granular gas is a large collection of mesoscopic or macroscopic
particles which collide \emph{inelastically} and are maintained in a
fluidized state. The prototypical model of a granular gas consists
of a dilute system of smooth inelastic hard spheres characterized by
a constant coefficient of normal restitution $\alpha<1$ \cite{BP04}.
A kinetic theory description based on the Boltzmann equation for
inelastic collisions has proven its relevance for the understanding
of the main properties of dilute granular gases, such as
non-equipartition of energy and high-velocity tails in uniform
states or non-Newtonian transport properties in non-uniform steady
states.

The kinetic theory definition of nonequilibrium temperature $T$ as a
measure of the mean kinetic energy (in the Lagrangian frame) per
particle can be directly extended to granular gases (except that it
is customary to measure $T$ in energy units, i.e., to take the
Boltzmann constant equal to unity). The key difference with respect
to normal gases is that the inelastic character of  collisions gives
rise to an energy sink, so that the temperature change per unit time
due to collisions is $\left.{\partial_t
T}\right|_{\text{coll}}=-\zeta T$, where  $\zeta$ is the so-called
cooling rate. It is  given (in the Maxwellian approximation) by
\beq
\zeta=\frac{d+2}{4d}(1-\alpha^2)\omega,\quad \omega\equiv
\frac{8\pi^{(d-1)/2}}{(d+2)\Gamma(d/2)}n\sigma^{d-1}\sqrt{\frac{T}{m}},
\label{0.1}
\eeq
where $d$ is the dimensionality of the system, $n$ is the number
density, $\sigma$ is the diameter of a sphere,  and $m$ is its mass.
Application of the Chapman--Enskog (CE) method to the inelastic
Boltzmann equation allows one to derive the Navier--Stokes (NS)
constitutive equations for  inelastic hard spheres \cite{BDKS98}. In
particular, Newton's law reads
\beq
P_{ij}=p\delta_{ij}-\eta_{\NS}\left(\frac{\partial u_i}{\partial
x_j}+ \frac{\partial u_j}{\partial x_i}-\frac{2}{d}\nabla\cdot{\bf
u}\delta_{ij}\right),
\label{1.1}
\eeq
where $P_{ij}$ is the pressure tensor, $p=(1/d)\text{Tr}{\sf P}=nT$
is the hydrostatic pressure,  ${\bf u}$ is the flow velocity, and
$\eta_{\text{NS}}$ is the NS shear viscosity. The latter quantity is
given, in the simplest Sonine approximation, by \cite{BDKS98} \beq
\eta_\NS=\frac{nT}{\nu+\zeta/2},\quad
\nu=\frac{1+\alpha}{2}\left[1-\frac{d-1}{2d}(1-\alpha)\right]\omega.
\label{0.2} \eeq

\begin{figure}
  \includegraphics[width=\columnwidth]{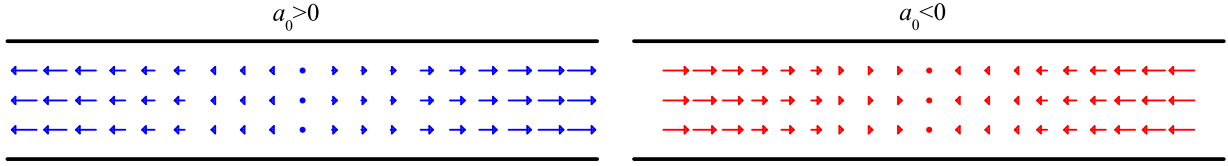}
\caption{Sketch of the uniform longitudinal flow for $a_0>0$ (left)
and $a_0<0$ (right).
\label{sketch}}
\end{figure}

The uniform longitudinal viscous flow is an unsteady
\emph{compressible} flow defined by a linear longitudinal velocity
field, a uniform  density, and a uniform  temperature \cite{GK96}.
The exact balance equations for mass and energy yield
\beq
u_x(x,t)=a(t)x, \quad n(t)=\frac{n_0}{a_0}a(t),\quad
a(t)=\frac{a_0}{1+a_0t},
\label{0.3}
\eeq
\beq
\partial_t T(t)=-\frac{2a_0}{dn_0}P_{xx}(t)-\zeta(t) T(t).
\label{0.4}
\eeq
This simple flow is also known as {\em
homo-energetic extension\/} and, along with the uniform shear flow
\cite{TM80,GS03}, is a particular case of a more general class of
homo-energetic affine flows characterized by $\partial^2
u_i/\partial x_j\partial x_k=0$ \cite{TM80}. The initial
longitudinal deformation rate rate $a_0$ is the only {\em control\/}
parameter determining the departure of the fluid from the
homogeneous state, thus playing a role similar to that of the shear
rate in the uniform shear flow state. On the other hand, in contrast
to the uniform shear flow, the sign of $a_0$ plays a relevant role
and defines two distinct situations (see Fig.\ \ref{sketch}). The
case $a_0>0$ corresponds to a progressively  slower {\em
expansion\/} of the gas from the plane $x=0$ into all of space.  On
the other hand, the case $a_0<0$ corresponds to a progressively
faster {\em compression\/} of the gas towards the plane $x=0$. The
latter takes place over a {\em finite\/} time period $t=|a_0|^{-1}$.
However, since the collision frequency rapidly increases with time,
the finite period $t=|a_0|^{-1}$ comprises an {\em infinite\/}
number of collisions per particle (see below).


According to Newton's law \eqref{1.1}, the normal stress $P_{xx}$ in
the uniform longitudinal flow is given by
\beq
T_x^*(t) \equiv
\frac{P_{xx}(t)}{p(t)}=1-2\frac{d-1}{d}\frac{a^*(t)}{1+\zeta^*/2},\quad
a^*(t)\equiv \frac{a(t)}{\nu(t)}\propto
\frac{a_0}{n_0}\frac{1}{\sqrt{T(t)}},
\label{0.5}
\eeq
where $\zeta^*\equiv \zeta/\nu$ is a dimensionless constant (except
that it depends on $d$ and $\alpha$). Note that $T_x^*$ is the ratio
between the ``longitudinal'' temperature $T_x=P_{xx}/n$ and the true
granular temperature $T$. The absolute value of the reduced
longitudinal rate $a^*$ plays the role of the Knudsen number of the
problem. The linear relationship \eqref{0.5} between $T_{x}^*$ and
$a^*$ is expected to hold for small values of $|a^*|$ only. More in
general, one can define a (reduced) nonlinear viscosity function
$\eta^*(a^*)$ by the relation
\beq
T_{x}^*(t)=1-2\frac{d-1}{d}\eta^*(a^*(t))a^*(t),\quad\lim_{a^*\to
0}\eta^*(a^*)=\frac{1}{1+\zeta^*/2}\equiv c_0.
\label{0.6}
\eeq
The  expansion of $\eta^*(a^*)$ in powers of $a^*$ yields the CE
series
\beq
\eta^*(a^*)=\sum_{k=0}^\infty c_k {a^*}^k,
\label{0.7}
\eeq
where
$c_1$ is a Burnett coefficient, $c_2$ is a super-Burnett
coefficient, and so on.

In this paper a kinetic theory description of the longitudinal
viscous flow defined by Eqs.\ \eqref{0.3} and \eqref{0.4} is
presented. As will be shown, a steady-state value of the temperature
is reached only if $a_0<0$. The main result is the  derivation of a
nonlinear first-order differential equation for $\eta^*(a^*)$. This
equation allows one to get the CE coefficients $c_k$ of Eq.\
\eqref{0.7} in a recursive way. The results show that the CE series
is convergent for inelastic hard spheres, while it diverges in the
elastic case. The viscosity function $\eta^*(a^*)$ for any value of
the inelasticity parameter $\alpha$ can be obtained by numerically
solving the differential equation with appropriate boundary
conditions. Finally,  a simplified analytical solution hardly
distinguishable from the numerical solution is proposed.

\section{Kinetic model}
In order to obtain explicit results, it is convenient to consider
the following BGK-like kinetic model \cite{BDS99} of the inelastic
Boltzmann equation:
\beq
(\partial_t
+\mathbf{v}\cdot\nabla)f=-\nu(f-f_\text{hcs})+\frac{\zeta}{2}\partial_\mathbf{v}\cdot[(\mathbf{v}-\mathbf{u})f],
\label{n1}
\eeq
where $f$ is the velocity distribution function and $f_\text{hcs}$
is the local version of the homogeneous cooling state distribution
{\cite{BP04}}. The NS viscosity coincides with that of the Boltzmann
equation if $\nu$ is given by Eq.\ \eqref{0.2}. However, a simpler
and more consistent choice is $\nu=\frac{1}{2}(1+\alpha)\omega$
\cite{SA05}, so that $\zeta^*=[(d+2)/2d](1-\alpha)$. In the uniform
longitudinal flow the velocity distribution function $f(x,{\bf
v},t)$ becomes spatially uniform when the velocities are referred to
a Lagrangian frame moving with the flow, i.e., $f(x,{\bf
v},t)=f({\bf V},t)$, where ${\bf V}\equiv {\bf v}-{\bf u}(x,t)$ is
the peculiar velocity. After simple algebra, Eq.\ \eqref{n1} can be
rewritten as
\beq
\left({\partial_\tau}-a_0{\partial_{V_x}}V_x\right)\overline{f}=-\overline{\nu}\left(\overline{f}-\overline{f}_\text{hcs}\right)+
\frac{\overline{\zeta}}{2}\partial_\mathbf{V}\cdot
\left(\mathbf{V}\overline{f}\right),
\label{2.3}
\eeq
where
\beq
\overline{f}({\bf V},\tau)\equiv\frac{n_0}{n(t)}f(x,{\bf v},t),\quad
\overline{\nu}(\tau)\equiv \frac{n_0}{n(t)}\nu(t),\quad
\overline{\zeta}(\tau)\equiv\frac{n_0}{n(t)}\zeta(t), \quad
\tau\equiv a_0^{-1}\ln(1+a_0t).
\label{2.4}
\eeq
Equation (\ref{2.3}) shows that the original uniform longitudinal
flow problem can be mapped onto the equivalent problem of a {\em
uniform\/} gas with a velocity distribution $\overline{f}$ and
subject to the action of a non-conservative force $-m a_0V_x
\widehat{\bf x}$. The density associated with $\overline{f}$ is the
initial density $n_0$ and thus remains constant. In fact,
$\overline{f}$ is the {\em probability\/} distribution function
normalized to $n_0$. The time variable $\tau=\int_0^t \dd t'
n(t')/n_0$ is a nonlinear measure of time scaled with the number
density. It is unbounded even if $a_0<0$ since in that case $\tau\to
\infty$ when $t\to |a_0|^{-1}$.

Taking velocity moments on both sides of Eq.\ \eqref{2.3} one gets
\beq
\partial_\tau T+\frac{2}{d}{a_0}{T}_{x}=-\overline{\zeta} T,\quad
\partial_\tau
{T}_{x}+2a_0{T}_{x}=-\overline{\nu}\left({T}_{x}-T\right)-\overline{\zeta}
{T}_{x}.
 \label{2.5}
\eeq
If $a_0>0$, $\partial_\tau T<0$ even in the elastic case
($\zeta=0$). As a consequence, the gas cools down and the reduced
longitudinal rate $a^*=a/\nu\sim 1/\sqrt{T}$ increases without
bound, i.e., $\lim_{\tau\to\infty}a^*(\tau)=\infty$. On the other
hand, in the compression case ($a_0<0$) the ``viscous heating'' term
$(2/d)|a_0|T_x$ competes with the ``inelastic cooling'' term
$\overline{\zeta}T$, so that the temperature either decreases or
increases with time depending on whether in the initial state one
has $\overline{\zeta}T>(2/d)|a_0|T_x$ or
$\overline{\zeta}T<(2/d)|a_0|T_x$, respectively. In either case, a
steady-state temperature $T_s$ is eventually reached when both
effects cancel each other. Therefore, if $a_0<0$ one has
$\lim_{\tau\to\infty}a^*(\tau)=a^*_s$.  The steady-state values for
the reduced longitudinal rate $a^*$, the temperature ratio $T_x/T$,
and the reduced viscosity $\eta^*$ can be easily obtained from Eq.\
\eqref{2.5} and are given by
\beq
a^*_s=-\frac{d}{2}\zeta^* \frac{1+\zeta^*}{1+d\zeta^*},\quad
T_{x,s}^*=\frac{1+d\zeta^*}{1+\zeta^*},\quad
\eta^*_s=\frac{1+d\zeta^*}{(1+\zeta^*)^2}.
\label{2.7}
\eeq
Comparison between $\eta^*_s$ and the NS value $c_0$ in Eq.\
\eqref{0.6} shows that the steady state is inherently non-Newtonian,
even in the quasi-elastic limit $\zeta^*\ll 1$, in which case
$c_0\approx 1-\zeta^*/2$ whilst $\eta_s^*\approx 1+(d-2)\zeta^*$.
Note that in the elastic case ($\zeta=0$) with $a_0<0$ only the
viscous heating term is present in the energy balance equation, so
that $\lim_{\tau\to\infty}a^*(\tau)=0$.

In summary,  the zero Knudsen number value ($a^*=0$) is a
``repeller'' of the time evolution of $a^*(\tau)$
 in the inelastic case. For elastic collisions,
the state $a^*=0$ is also a repeller if $a_0>$ but it is an
``attractor'' of $a^*(\tau)$ if $a_0<0$. Figure \ref{fig}(a) shows
$a_s^*$, $T_{x,s}^*$, and $\eta_s^*$ as functions of $\alpha$ for
three-dimensional systems ($d=3$). It is interesting to note that
the steady-state reduced viscosity $\eta^*_s$ is nearly independent
of $\alpha$.

As noted before, if $a_0<0$ the whole compression process takes
place during a finite time period $t=|a_0|^{-1}$.
 Let us see now that the corresponding number of collisions per particle is infinite.
The number of collisions per particle during an elementary
time interval $\dd t$ is $\dd s=\nu\dd t=\overline{\nu}\dd\tau$. The
accumulated number of collisions between the initial state and the
scaled time $\tau$ is then $s(\tau)=\nu_0\int_0^\tau\dd \tau'
\sqrt{T(\tau')/T_0}$, where $\nu_0$ and $T_0$ are the initial
collision frequency and temperature, respectively. If $a_0<0$ the
temperature is bounded between $T_0$ and $T_s$, i.e.,
$\text{min}(T_0,T_s)\leq T(\tau')\leq \text{max}(T_0,T_s)$.
Therefore, $\text{min}(1,\sqrt{T_s/T_0})\nu_0\tau\leq s(\tau)\leq
\text{max}(1,\sqrt{T_s/T_0})\nu_0\tau$. Since $\tau\to\infty$ when
$t\to|a_0|^{-1}$, it follows that $s\to\infty$ in that limit.

\section{Nonlinear viscosity}
\begin{figure}
  \includegraphics[width=\columnwidth]{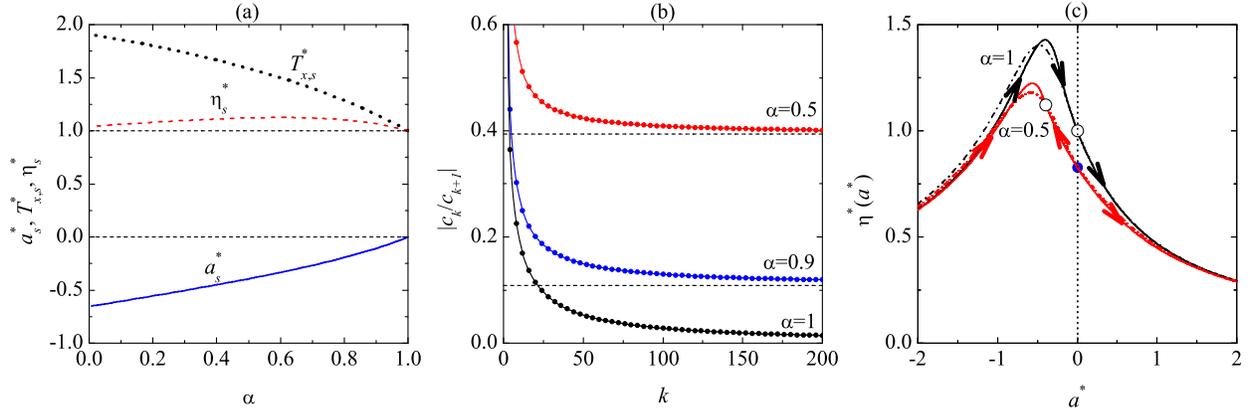}
\caption{(a) Plot of the steady-sate values $a_s^*$, $T_{x,s}^*$, and $\eta_s^*$ as functions of $\alpha$. (b) Ratio $|c_k/c_{k+1}|$
for $\alpha=0.5$, $0.9$, and $1$; the horizontal dashed lines indicate the corresponding values of $|a_s^*|$. (c) Nonlinear viscosity for
$\alpha=0.5$ and 1 as obtained from the numerical solution of Eq.\ \eqref{3.2} (solid lines) and from Eq.\ \eqref{3.7} (dash-dotted lines); the circles
denote the steady-state points and the arrows indicate the direction of the time evolution.
All the panels correspond to three-dimensional hard spheres ($q=\frac{1}{2}$, $d=3$).
\label{fig}}
\end{figure}

In order to get the whole rheological function $\eta^*(a^*)$ we need
to eliminate time in favor of $a^*(\tau)=a_0/\overline{\nu}(\tau)$,
taking into account that both $\overline{\nu}$ and
$\overline{\zeta}$ are proportional to $[T(\tau)]^q$, where
$q=\frac{1}{2}$. It is convenient to consider for the moment $q$ as
an arbitrary positive parameter and set $q=\frac{1}{2}$ at the end.
Thus, since $T_x=T_x^*T$, one has
\beq
\partial_\tau T_x=T_x^*\partial_\tau T+T(\partial_\tau a^*)\partial_{a^*}T_x^*=\left(T_x^*-qa^*\partial_{a^*}T_x^*\right)\partial_\tau T.
\label{3.1}
\eeq
Making use of Eq.\ \eqref{2.5} it is easy to get the following
nonlinear first-order differential equation for the function
$\eta^*(a^*)$:
\begin{equation}
\eta^*-\left(1-2\frac{d-1}{d}\eta^*
a^*\right)\left(1+\frac{2}{d}\eta^*
a^*\right)+q\left[\zeta^*+\frac{2}{d}a^*\left(1-2\frac{d-1}{d}\eta^*
a^*\right)\right]\left(\eta^*+a^*\partial_{a^*}\eta^*\right)=0.
\label{3.2}
\end{equation}
Inserting the CE expansion \eqref{0.7} one can get the coefficients
$c_k$ in a recursive way:
\beq
c_0=\frac{1}{1+q\zeta^*},\quad
c_k=-\frac{2}{d}\frac{1}{1+(k+1)q\zeta^*}\left\{(d-2+kq)c_{k-1}+2\frac{d-1}{d}\sum_{\ell=0}^{k-2}c_\ell
c_{k-2-\ell} \left[1-(\ell+1)q\right]\right\} .
\label{3.3}
\eeq
Figure \ref{fig}(b) shows the absolute value of the ratio
$c_k/c_{k+1}$ for three-dimensional hard spheres ($q=\frac{1}{2}$,
$d=3$) with $\alpha=0.5$ (highly inelastic case), $\alpha=0.9$
(moderately inelastic case), and $\alpha=1$ (elastic case). In the
inelastic cases the ratio $|c_k/c_{k+1}|$ tends to a non-zero finite
value, thus indicating that the CE series \eqref{0.7} converges. In
fact, the results are consistent with
$\lim_{k\to\infty}|c_k/c_{k+1}|=|a_s^*|$, i.e., the radius of
convergence of the CE series coincides with the steady-state value
of the Knudsen number. In contrast, in the elastic case one has
$\lim_{k\to\infty}|c_k/c_{k+1}|=0$, so that the radius of
convergence shrinks to zero and the CE series \eqref{3.3} diverges.
In fact, if $\zeta^*=0$, the magnitude of the coefficients $c_k$
grows so fast that the summation term in Eq.\ \eqref{3.2} can be
neglected and one has $|c_k/c_{k+1}|\approx (d/2q)k^{-1}$.

Inspection of Eq.\ \eqref{3.2} shows the following asymptotic
behaviors in the limit of $|a^*|\to\infty$:
\beq
\begin{array}{ll}
a_0>0:& \lim_{a^*\to\infty}\eta^*(a^*)a^*=\frac{d}{2(d-1)},\quad
\lim_{a^*\to\infty}T_x^*(a^*)=0,\\ a_0<0:&
\lim_{a^*\to-\infty}\eta^*(a^*)a^*=-\frac{d}{2},\quad
\lim_{a^*\to-\infty}T_x^*(a^*)=d.
\end{array}
\label{3.4}
\eeq
 Therefore, in the limit $a^*\to\infty$ there is no motion
(relative to the flow velocity) along the $x$ direction, while in
the opposite limit $a^*\to-\infty$ the motion becomes
one-dimensional along the $x$ axis.

\begin{figure}
  \includegraphics[height=.3\textheight]{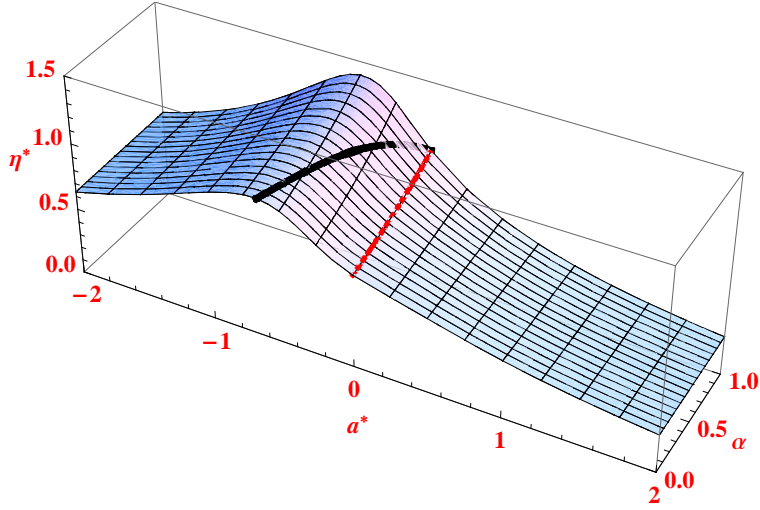}
\caption{Surface plot showing the dependence on both $a^*$ and $\alpha$ of of the viscosity function $\eta^*$ given by Eq.\ \protect\eqref{3.7}.
The thin and thick lines refer to the NS and steady-state points, respectively. \label{3D}}
\end{figure}

 The complete viscosity function
$\eta^*(a^*)$ for any $a^*$, including values of the Knudsen number
$|a^*|$ beyond the radius of convergence $|a^*_s|$, is obtained by
numerically solving the differential equation \eqref{3.2} with
appropriate boundary conditions. To get the branches $a^*>0$ and
$a_s^*<a^*<0$ one must start from a small value of $|a^*|$ and apply
the initial condition $\eta^*=c_0+c_1a^*$. The branch $a^*<a_s^*<0$
is obtained by starting from a large value of $a^*$ and applying the
boundary condition $\eta^*=-d/2a^*$. The resulting curve
$\eta^*(a^*)$ is plotted in Fig.\ \ref{fig}(c) for three-dimensional
hard spheres ($q=\frac{1}{2}$, $d=3$) with $\alpha=0.5$ and $1$. The
arrows on the curves indicate the direction followed by the time
evolution of $a^*(\tau)$.

While the differential equation \eqref{3.2} can be solved
numerically, it is interesting to approximate its solution by an
analytical expression. To that end, let us see $q$ as a small
perturbation parameter and formally expand the solution to Eq.\
\eqref{3.2} in powers of $q$: \beq
\eta^*(a^*)=\eta_0^*(a^*)\left[1-qh_1(a^*)+\cdots\right].
\label{3.5} \eeq By inserting Eq.\ \eqref{3.5} into Eq.\ \eqref{3.2}
and keeping terms linear in $q$, one gets, after some algebra,
\beq
\eta_0^*(a^*)=\left[\frac{1}{2}+\frac{d-2}{d}a^*+\beta(a^*)\right]^{-1},
\quad
h_1(a^*)=\frac{\zeta^*+\frac{1}{2}+a^*-\beta(a^*)}{4\beta^2(a^*)},\quad
\beta(a^*)\equiv
\sqrt{\left(\frac{1}{2}+a^*\right)^2-\frac{2}{d}a^*}.
\label{3.6}
\eeq
Note that the zeroth-order function $\eta_0^*(a^*)$ is
independent of the reduced cooling rate $\zeta^*$. Instead of
proceeding to higher order terms in the $q$-expansion \eqref{3.5},
let us consider the simple Pad\'e approximant
\beq
\eta^*(a^*)\simeq
\frac{\eta_0^*(a^*)}{1+qh_1(a^*)}.
\label{3.7}
\eeq
This approximation has the merit of providing correctly (for any
value of $q$) the values of the NS coefficient  $c_0$ [see Eq.\
\eqref{3.3}] and the steady-state point $\eta^*_s=\eta^*(a^*_s)$
[see Eq.\ \eqref{2.7}], as well as the limiting behaviors
\eqref{3.4}. Moreover, even in the physical case of hard spheres
($q=\frac{1}{2}$), Eq.\ \eqref{3.7} represents an excellent
analytical approximation to the numerical solution of Eq.\
\eqref{3.2}, especially for highly inelastic systems, as shown in
Fig.\ \ref{fig}(c). The discrepancies are only apparent near the
maximum of $\eta^*(a^*)$ and in the decay to the left of the
maximum. Figure \ref{3D} shows a surface plot of Eq.\ \eqref{3.7}.

\section{Conclusions}
The uniform longitudinal viscous flow is an unsteady compressible
flow [see Eqs.\ \eqref{0.3} and \eqref{0.4}] that, despite its
apparent simplicity, constitutes a non-trivial playground for
nonequilibrium statistical mechanics beyond the NS description.
While this state has been studied in the past for normal gases
 \cite{GK96}, the present work addresses the problem in the case of a
granular gas of inelastic hard spheres.  The relevant control
parameter is the reduced longitudinal rate $a^*(t)=a(t)/\nu(t)$
(whose magnitude plays the role of the Knudsen number of the
problem) and the relevant response function is the  time-dependent
generalized viscosity $\eta^*(a^*)$ defined by Eq.\ \eqref{0.6}. The
exact energy balance equation \eqref{0.4} shows that if $a_0>0$
(expansion states), the temperature monotonically decreases and so
$a^*(t)$ increases with time, for both  elastic ($\zeta=0$) and
inelastic ($\zeta>0$) collisions. On the other hand, if $a_0<0$
(compression states), $a^*(t)$ goes to zero from below in the
elastic case, whereas it tends to a stationary value $a_s^*<0$ in
the inelastic case [see Fig.\ \ref{fig}(a)].  Expressed in other
terms, one can say that the homogeneous state  ($a_0=0$ or $a^*=0$)
is \emph{unstable} against any  perturbation $a_0\neq 0$, no matter
how weak it is, if the particles are \emph{inelastic}. On the other
hand, $a^*=0$ is \emph{stable} for \emph{elastic} collisions if
$a_0<0$. Given that the CE expansion \eqref{0.7} is carried out
about (and measures the departure from) the reference homogeneous
state ($a^*=0$), it follows the ``arrow of time'' for granular gases
but goes against it for normal gases (if $a_0<0$). As a consequence,
it can be expected on physical grounds that the CE expansion
diverges in the case of elastic collisions but does converge if the
collisions are inelastic, the radius of convergence $|a_s^*|$
increasing with inelasticity.

To confirm the above expectation and also to get the rheological
function $\eta^*(a^*)$, the simple model kinetic equation \eqref{n1}
has been considered in this paper. The viscosity function
$\eta^*(a^*)$ predicted by the model is the solution to the
nonlinear first-order  differential equation \eqref{3.2} (with
$q=\frac{1}{2}$). The CE coefficients $c_k$ are shown to verify the
asymptotic law $\lim_{k\to\infty}|c_k/c_{k+1}|=|a_s^*|$, thus
confirming the convergence of the series \eqref{0.7} with a finite
radius $a^*=|a_s^*|$, which vanishes in the elastic limit [see Fig.\
\ref{fig}(b)]. The function $\eta^*(a^*)$ exhibits an interesting
behavior [see Fig.\ \ref{fig}(c)]. Starting from the NS value
$\eta^*(0)=c_0$, it decreases as the longitudinal rate $a^*$
increases, behaving as $\eta^*(a^*)\approx [d/2(d-1)]/a^*$ in the
limit of large positive $a^*$. In the latter limit, the gas becomes
highly anisotropic with no random motion along the flow direction.
The situation is even more interesting in the case of negative
$a^*$. As the magnitude of the longitudinal rate increases, the
viscosity function starts increasing, reaches a maximum value past
the steady-state point, and decreases thereafter, behaving as
$\eta^*(a^*)\approx (d/2)/|a^*|$ in the limit of large $|a^*|$. The
gas becomes highly anisotropic also in this limit, but this time
because the degrees of freedom orthogonal to the flow direction are
suppressed. As Fig.\ \ref{fig}(c) shows, the largest quantitative
influence of inelasticity on the shape of $\eta^*(a^*)$ occurs near
the maximum. It is also in this region where the analytical
approximation \eqref{3.7} exhibits the largest deviations from the
numerical solution of Eq.\ \eqref{3.2}. Otherwise, Eq.\ \eqref{3.7}
succeeds in capturing the main features of the solution, including
the steady-state and NS values, as well as the asymptotic behaviors.
However, an important property not accounted for by Eq.\ \eqref{3.2}
is the divergence of the CE series in the elastic case
($\zeta^*=0$). This is a consequence of the fact that Eq.\
\eqref{3.2} is based on the $q$-expansion \eqref{3.5} (truncated
after the first term) and the CE expansion in the reference system
($q=0$) is convergent, regardless the value of $\zeta^*$.

As a final point, it is worthwhile rewriting the CE expansion
\eqref{0.7} in the dimensional form
\beq
P_{xx}=p-2\frac{d-1}{d}\eta_\NS\sum_{k=0}^\infty
\frac{c_k}{\nu^kc_0}\left(\frac{\partial u_x}{\partial
x}\right)^{k+1}.
\label{4.1}
\eeq
When arbitrary hydrodynamic gradients are present, Eq.\ \eqref{4.1}
represents a \emph{partial} contribution to the general CE expansion
of the normal stress $P_{xx}$ . The advantage of the uniform
longitudinal flow is that it possesses an only hydrodynamic
gradient, $\partial u_x/\partial x$, which is spatially uniform.
Therefore, in this case the general CE expansion of $P_{xx}$ reduces
to Eq.\ \eqref{4.1}. However, the convergence or divergence of the
partial series \eqref{4.1} does not depend on whether the system is
actually  in the uniform longitudinal flow or in any other state.
According to the results derived in this paper, it turns out that
the partial series \eqref{4.1} diverges for normal gases but
converges for granular gases. A similar conclusion is reached in the
case of the expansion of the shear stress $P_{xy}$ in powers of the
shear rate $\partial u_x/\partial y$ \cite{S08}.


\begin{theacknowledgments}
 This work has been supported by the
Ministerio de Educaci\'on y Ciencia (Spain) through Grant No.\
FIS2007--60977 (partially financed by FEDER funds) and by the Junta
de Extremadura (Spain) through Grant No.\ GRU08069.
\end{theacknowledgments}



\bibliographystyle{aipproc}   

\end{document}